\documentclass[11pt,showpacs]{revtex4-1}

\usepackage{epsfig}
\input epsf.tex
\usepackage{amssymb}
\usepackage{amsmath}
\usepackage{amsthm}
\usepackage{amsopn}

\def\comment#1{}

\def\beq{\begin{equation}}
\def\eeq{\end{equation}}
\def\bea{\begin{eqnarray}}
\def\eea{\end{eqnarray}}

\usepackage{ulem}
\usepackage{cancel}
\usepackage{color}

\begin{document}

\title{  The B-mode polarization of CMB and Cosmic Neutrino Background  }

\author{Rohoollah Mohammadi$^{1}$\footnote{mohammadi0692@gmail.com},  Jafar Khodagholizadeh $^{2}$\footnote{gholizadeh@ipm.ir},  M. Sadegh$^{3}$ and She-Sheng Xue$^{4}$\footnote{xue@icra.it}}
\affiliation{$^{1}$ Iran Science and Technology Museum (IRSTM), PO BOX: 11369-14611, Tehran, Iran.\\
$^{2}$ Shahid Beheshti Bracnch, Farhangian University, Tehran, Iran.\\
 $^{3}$  Department of Physics, Payame Noor University (PNU) , P. O. Box: 19359-3697, Tehran, Iran \\
$^{4}$ ICRANet, P.zza della Repubblica 10, I--65122 Pescara, Physics Department, University of Rome {\it La Sapienza}, P.le Aldo Moro 5, I--00185 Rome, Italy.}

\begin{abstract}
It is known that in contrast with the E-mode polarization, the B-mode polarization of the Cosmic Microwave Background cannot be generated by the Compton scattering in the case of scalar mode of metric perturbation. However it is possible to generate the B-mode by the  Compton scattering in the case of tensor mode of metric perturbation. For this reason, the ratio of tensor to scalar modes of metric perturbation ($r\sim C_{Bl}/C_{El}$) is estimated by comparing the B-mode power spectrum with the E-mode at least for small $l$.
We study the CMB polarization specially B-mode due to the weak interaction of Cosmic Neutrino Background (CNB) and CMB, in addition to the Compton scattering in both cases of scalar and tensor metric perturbations.
It is shown that the power spectrum $C_{Bl}$ of the B-mode polarization receives some contributions from scalar and tensor modes, which have  effects on the value of
$r$-parameter. We also show that the B-mode polarization power spectrum can be used as an indirect probe into the CNB.
\end{abstract}


\maketitle

\section{Introduction}
The tensor and scalar metric perturbations are described by their power spectra
$P_{T}=A_{T} (\frac{k}{k})^{n_{T}-1}$ and $P_{s}=A_{T} (\frac{k}{k})^{n_{s}-1}$,
$n_{T,s}$ and $A_{T,s}$ are their spectral indices and amplitudes.
The tensor-to-scalar ratio $r=P_{T}/P_{S}$ and its value
are very important to verify or constrain the standard scenario of Big-Bang and inflation model. Comparing the curl-like mode (B-mode) with divergence-like (E-mode) in
linearly polarized CMB power spectra is one of the most important
methods measuring the r-parameter value. The
detection of the B-mode signals induced by the primordial  gravitational  waves
can be  one  of  the  main  targets  in
many ongoing and future CMB experiments \cite{nagata}.
The BICEP2 collaboration detected the
B-mode in the primordial cosmic microwave background (CMB) for the first time
and announced  $r=0.20_{-0.05}^{+0.07}$ $( 68\% CL) $ \cite{BICEP2}.
This result is not consistent with the Planck limit, $r<0.11 ( 98 \% CL)$.
Some authors speculated that the observed B-mode polarization is the result
of a primordial Faraday rotation of the E-mode polarization \cite{14,Giovannini}.
The BICEP2 data could be explained by the scalar and tensor modes from primordial magnetic fields \cite{Bovin}. The interaction between CMB photons and CNB Neutrinos can produce
the B-mode spectrum \cite{Khodagholizadeh} even with the scalar perturbation only.
The Planck group presented discussions that the result of BICEP2 can be fully
attributed to cosmic dust \cite{dust}. The recent
Bicep/Keck Array observation reported upper bounds on the tensor-to-scalar
ratio, $r_{0.05}<0.09$ and $r_{0.05}<0.07$ at $( 95\%)$ C.L. by using B-modes alone and combining the B-mode results
with Planck temperature analysis, respectively \cite{keck}.
It is crucial to know all possible effects and interactions can affect
the r-parameter value. Therefore, continuing our previous work \cite{Khodagholizadeh},
we study the B-mode spectrum by considering the CMB interacting with the CNB
in both cases of scalar and tensor metric perturbations in this article.

\comment{
The rate of photon-neutrino scattering is very small due to the smallness of photon-neutrino cross-section (in order of $(\alpha G^{F} m_{\nu})^{2}$ ) 
\cite{Dicus}.
As discussed in Ref.~\cite{r.mohammadi,Khodagholizadeh}, the time evaluation of density-matrix
elements of CMB $\rho_{ij}$, i.e., Stokes parameters, can be simply written as:
\begin{equation}
\frac{d \rho_{ij}}{dt}=<[H^{0},\rho_{ij}]>-\frac{1}{2}\int dt <[H^{0},[H^{0},\rho_{ij}]]>
\end{equation}
where $H^{0}$ is the first order of magnitude of photon-neutrino effective Hamiltonian
and  $<[H^{0},\rho_{ij}]>$ is the forward scattering amplitude which is of the order
of $ H^{0}$ or $(\alpha G^{F} m_{\nu})$ instead of  $<[H^{0},[H^{o},\rho_{ij}]]>$ or $(\alpha G^{F})^{2}$ for photon-neutrino cross-section. As shown in \cite{r.mohammadi,Khodagholizadeh}, the rate of generation of circular and linear polarizations due to photon-neutrino interaction depends on forward scattering, changing polarization without changing momentum, or $(\alpha G^{F})$ which is very larger than the rate of photon-neutrino scattering due to photon-neutrino cross-section,  changing polarization and changing momentum too.
In the standard scenario of Big-Bang, neutrinos are decoupled from the photon-baryon fluid
at around $T=1$ MeV long before photon decoupling. But this doesn't means photon-neutrino cannot interact.}

\section{The generation of polarized CMB via scattering.}

Polarized CMB photons are described by the density and number operators
\bea
\hat\rho=\frac{1}{\rm {tr}(\hat \rho)}\int\frac{d^3p}{(2\pi)^3}
\rho_{ij}(p)D_{ij}(p),
\eea
where $\rho_{ij}(p)$ represents the general density-matrix in the space of
polarization states with a fixed energy-momentum ``$p$'', correspondingly
$D_{ij}(p)$ is the number operator whose expectation value
\bea
\langle\, D_{ij}(p)\,\rangle\equiv {\rm tr}[\hat\rho
D_{ij}(p)]=(2\pi)^3 \delta^3(0)(2p^0)\rho_{ij}(p).\label{t1}
\eea
The quantum Boltzmann equation for the time-evolution of the density matrix
of the polarized CMB photons reads as,
\begin{equation}\label{qbe}
\frac{d \rho_{ij}}{dt}=<[H^{0},\rho_{ij}]>-\frac{1}{2}\int dt <[H^{0},[H^{0},\rho_{ij}]]>
\end{equation}
where $H^{0}$ is the effective Hamiltonian. The first term in the RHS is the forward scattering amplitude and the second one represents the high-order collision terms.
Equation (\ref{qbe}) can be written as \cite{cosowsky1994},
\bea
(2\pi)^3 \delta^3(0)(2p^0)
\frac{d}{dt}\rho_{ij}(p) = i\langle \left[H_I
(t);D_{ij}(p)\right]\rangle-\frac{1}{2}\int dt\langle
\left[H_I(t);\left[H_I
(0);D_{ij}(p)\right]\right]\label{bo}\rangle,
\eea
and $H_I(t)$ is the first order of the interacting Hamiltonian. The LHS of this equation is known as the Liouville term dealing with the effects of gravitational perturbations around the homogeneous cosmology. Using this equation, many authors investigated  the effects of the Compton scattering (electromagnetic interactions) on the anisotropy and polarization of CMB (see for example \cite{cosowsky1994,zal,hu}). We have studied in \cite{Khodagholizadeh} the
effects of the scattering between CMB photons and CNB neutrinos in the case of scalar metric perturbation.

We adopt the Stokes parameters I, Q, U and V to describe the polarizations of CMB photons,
\bea
\hat{\rho}=\left(\begin{array}{cc}
             I+Q& U-iV \\
             U+iV & I-Q \\
                 \end{array}
        \right),
\label{i-v} \eea
where the parameter $I$
is total intensity, $Q$ and $U$ are intensities of linear
polarizations, whereas the parameter $V$
indicates net circular polarization or
the difference between left- and right-circular polarizations
intensities.
The time evaluation of stocks parameters for linear polarization
with collision terms on the right side of Boltzmann equation (\ref{bo}), include the Compton scattering and photon-neutrino interaction \cite{Khodagholizadeh},
\begin{eqnarray}
\frac{d}{dt}(Q\pm iU)  &\approx & C^\pm_{e\gamma} \mp i \dot{\kappa}_{\pm} (Q\pm iU),
\label{bo1}
\end{eqnarray}
where RHS  of the first term $C^\pm_{e\gamma}$ comes from the effects of Compton scattering (see for example \cite{cosowsky1994,zal,hu}),
whereas the second term comes from the photon-neutrino interaction and
$\dot{\kappa}_{\pm}$ is given by
\begin{eqnarray}
\dot{\kappa}_{\pm} &=& -\frac{\sqrt{2}}{6\pi k^0}\alpha\,G^F\int d\mathbf{q} f_\nu(q)
\times
\varepsilon_{\mu\nu\rho\sigma}\,\epsilon_2^{\mu}\epsilon_{1}^\nu k^\rho q^\sigma
\nonumber\\
&=&  \frac{\sqrt{2}}{6\pi k^0}\alpha\,G^F\int d\mathbf{q} f_\nu(q)
\times \left[q^0 \vec{k}.(\epsilon_1\times\epsilon_2)+k^0 \vec{q}.(\epsilon_1\times\epsilon_2)\right]\label{kappa}\\
&=&  \frac{\sqrt{2}}{6\pi}\alpha\,G^F\left[\frac{n_\nu}{2}+\int d\mathbf{q} f_\nu(q)
\vec{q}.(\epsilon_1\times\epsilon_2)\right],
\label{kappa1}
\end{eqnarray}
which is expressed in terms of the CNB neutrino number density $n_\nu$ and bulk velocity
\begin{eqnarray}
  q^i &=& \frac{1}{n_\nu}\int d\mathbf{q}q^i\,f_\nu. \label{average}
\end{eqnarray}
If we apply above equation for Cosmic Neutrino Background (CNB), the second term on the right is much smaller than the first one due to the
smallness of the CNB neutrino bulk velocity. For the rest of paper, we consider
$\dot{\kappa}_{\pm}\simeq  \frac{\sqrt{2}}{12\pi}\alpha\,G^F\,n_\nu$.
Note that the time evolution of CMB circular polarization $V$- Stokes parameter
was investigated in Ref.~\cite{r.mohammadi}, and the total density
$I$ does not vary in time.

\section{Power Spectrum of Scalar Modes}
As shown in \cite{cosowsky1994,zal}, the Compton scattering in the  presence of scalar perturbation can be a source for the E-mode of the CMB linear polarization.
The B-mode one can only be generated by the Compton scattering in the presence of tensor perturbation. However, the B-mode polarization of CMB in scalar perturbation can be generated by the CMB-CNB interaction together with the Compton scattering \cite{Khodagholizadeh}, and
we generalize these results to the case of tensor perturbations
in the following.  The scalar and tensor perturbations of metric are indicated by $S$ and  $T$ superscript respectively. In general, for a given mode of metric perturbation $\mathbf{K}$,
we can choose the coordinate system where
$\mathbf{K} \parallel \hat{\mathbf{z}}$ and
$(\hat{\mathbf{e}}_1,\hat{\mathbf{e}}_2)=(\hat{\mathbf{e}}_\theta,
\hat{\mathbf{e}}_\phi)$. Also for each plane wave, we describe the scattering
as the transport through a plane parallel medium \cite{chandra,kaiser}.
In the case of the scalar perturbation, the Boltzmann equation (\ref{bo1})
is given as \cite{Khodagholizadeh},
\begin{eqnarray}
\frac{d}{d\eta}\Delta_I^{(S)} +iK\mu \Delta_I^{(S)}+4[\dot{\psi}-iK\mu \phi]
&=&\dot\tau[-\Delta_I^{(S)} +
\Delta_{I0}^{(S)} +i\mu v_b +{1\over 2}P_2(\mu)\Pi]  \label{Boltzmann}\\
\frac{d}{d\eta}(Q^{(S)}\pm iU^{(S)}) +iK\mu (Q^{(S)}\pm iU^{(S)}) &=& \dot\tau[
-(Q^{(S)}\pm iU^{(S)}) -{1\over 2} [1-P_2(\mu)] \Pi]\nonumber\\&\mp& i\,a(\eta) \,\dot{\kappa}_{\pm}\,(Q^{(S)}\pm iU^{(S)})
\label{Boltzmann1} \\
\Pi&=&\Delta_{I2}^{(S)}
+\Delta_{P2}^{(S)}+
\Delta_{P0}^{(S)}, \nonumber
\end{eqnarray}
where $\eta $ is the conformal time.
The differential optical depth  for Compton scattering is denoted by $\dot{\tau}=an_ex_e\sigma_T$, where $a(\eta)$ is the expansion factor normalized
to unity today, $v_b$ is baryon bulk velocity, $\mu=\hat{\mathbf{K}}\cdot \hat{n}$ is the angle between the CMB photon direction $\hat{n}$ and wave number $\mathbf{K}$,  $n_e$ is the electron number density, $x_e$ is the ionization
fraction and  the Thomson cross section is indicated by $\sigma_T$. The second term in R.H.S of above equation is the contribution of photon-neutrino forward scattering  with $\dot{\kappa}_{\pm}$ coefficients.
The source terms in these equations involve
the multipole moments of temperature and polarization which
are defined as $ \Delta(K,\mu)=\sum_l(2l+1)(-i)^{l}\Delta_l(k)P_l(\mu)$,
where $P_l(\mu)$ is the Legendre polynomial of order $l$.
Temperature anisotropies $\Delta_I^{S}$ respect to $\Delta_P^S$ have additional sources
in metric perturbations $\phi$ and $\psi$
and in baryon velocity term $v_b$. By defining $\Delta _{P}^{\pm (S)}=Q^{(S)}\pm iU^{(S)}$, we rewrite Eq.(\ref{Boltzmann1}) as
following,
\begin{equation}\label{Boltzmann2}
    \frac{d}{d\eta}\left[\Delta _{P}^{\pm (S)}\,e^{iK\mu\eta\, \pm\, i\tilde{\kappa}(\eta,\mu)\,+\tilde{\tau}(\eta)}\right]  = -e^{iK\mu\eta\, \pm\, i\tilde{\kappa}(\eta)\,+\tilde{\tau}(\eta)}\left(
{1\over 2} \dot\tau[1-P_2(\mu)] \Pi\right),
\end{equation}
where
\begin{equation}\label{tilde}
    \tilde{\kappa}(\eta,\mu)=\int_0^{\eta}\, d\eta\, a(\eta)\,\dot{\kappa}_{\pm},\,\,\,\,\,\,\tilde{\tau}(\eta)=\int_0^{\eta}\, d\eta\, \dot{\tau}.
\end{equation}
 To obtain the value of $\Delta _{P}^{\pm (S)}(\hat{n})$ at present time $\eta_0$ and direction $\hat{n}$, in addition to
integrate the Boltzmann equation (\ref{Boltzmann1})
along the line of sight \cite{sz}, one needs to evolve the anisotropies until the present epoch and integrate over all the Fourier modes $K$,
\begin{eqnarray}
\Delta _{P}^{\pm (S)}(\hat{\bf{n}})
&=&\int d^3 \bf{K} \xi(\bf{K})e^{\pm2i\phi_{K,n}}\Delta _{P}^{\pm (S)}
(\eta_0,K,\mu),\,\,\,\,\,\kappa(\eta)=\int_{\eta}^{\eta_0}\, d\eta\, a(\eta)\,\dot{\kappa}_{\pm}(\eta)\nonumber\\
 \Delta _{P}^{\pm (S)}
(\eta_0,K,\mu)&=&{3 \over 4}(1-\mu^2)\int_0^{\eta_0} d\eta\,
e^{ix \mu \pm  i\kappa(\eta) -\tau}\,\,\Pi(K,\eta)\label{Boltzmann3}
\end{eqnarray}
where $x=K(\eta_0 - \eta)$,
$\phi_{\bf K,n}$ is the angle needed to rotate
the $\bf{K}$ and $\hat{\bf{n}}$ dependent basis to a
fixed frame in the Sky, and $\xi(\bf{K})$
is a random variable used to characterize the initial
amplitude of the mode  $\bf{K}$ which follows the statistical property
\begin{equation}\label{xi}
    \langle \xi^{*}({\bf K}_1)\xi({\bf K}_2)
\rangle= P_{S}({\bf K})\delta({\bf K}_1- {\bf K}_2)
\end{equation}
which $P_S(K)$ is the initial power spectrum of scalar perturbation. One can separate
the CMB  polarization  in terms of  the curl-free part (E-mode)
\begin{equation}\label{Emode}
  \Delta_{E}^{(S)}(\hat{n})\equiv-\frac{1}{2}[\bar{\eth}^{2}\Delta_{P}^{+(S)}(\hat{\bf{n}})+\eth^{2}\Delta_{P}^{-(S)}(\hat{\bf{n}})]
\end{equation}
 and divergence-free part (B-mode)
 \begin{equation}\label{Bmode}
    \Delta_{B}^{(S)}(\hat{n})\equiv\frac{i}{2}[\bar{\eth}^{2}\Delta_{P}^{+(S)}(\hat{\bf{n}})-\eth^{2}\Delta_{P}^{-(S)}(\hat{\bf{n}})]
 \end{equation}
 that $\eth$ and $\bar{\eth}$ are spin raising and lowering operators respectively. By assuming that scalar metric perturbations are axially-symmetric around ${\bf K}$ so that $\eth^{2}=\bar{\eth}^{2}=\partial_{\mu}^{2}$, the E- and B-modes are given as
\begin{eqnarray}
\Delta_{E}^{(S)}(K,\eta_0)&=&
-\int_{0}^{\eta_{0}}d\eta g(\eta)\frac{3}{4}\Pi(K,\eta)\partial_{\mu}^{2} \left[(1-\mu^{2})^2e^{ix\mu} \cos{\kappa(\eta)}\right], \label{Emode1}\\
\Delta_{B}^{(S)}(K,\eta_0)&=&\int_{0}^{\eta_{0}}d\eta g(\eta)\frac{3}{4}\Pi(K,\eta)\partial_{\mu}^{2} \left[ (1-\mu^{2})^2e^{ix\mu} \sin{\kappa(\eta)}\right],\label{Bmode1}
\end{eqnarray}
where $g(\eta)=\dot{\tau}e^{-\tau}$. Note that the Compton scattering can not generate
B-mode without taking into account tensor type of metric perturbations \cite{zh95,uros,letter,hu}, and we check this in our calculations.
The polarized spectrum of CMB is then obtained by integrating over the initial power spectrum of the metric perturbation. As a result, the power spectrum for $E$ and $B$ modes are given by
\begin{equation}\label{CBS1}
    C_{E,B l}^{(S)}=\frac{1}{2l+1}\frac{(l-2)!}{(l+2)!}\int d^3K P_S(K)|\sum_m\int d\Omega Y^*_{lm}\Delta_{E,B}^{(S)}|^2.
\end{equation}
By taking the photon-neutrino scattering into account, we showed \cite{Khodagholizadeh}
the $C^{(S)}_{Bl}$ is not zero for the scalar perturbation.
The CMB polarized power spectra in multipole moments $l$ are $C_{E l}^{(S)}=\bar{C}^{(S)}_{E,l} \, (\cos^2{\bar{\kappa}})$ and $C_{B l}^{(S)}=\bar{C}^{(S)}_{E,l} \, (\sin^2{\bar{\kappa}})$. $\bar{C}^{(S)}_{E,l}$ is the value of E-mode polarized power spectrum attributed to
the Compton scattering in the case of scalar perturbation \cite{zal}:
 \begin{eqnarray}
 \bar{C}^{(S)}_{E,l}&=&(4\pi)^2\frac{(l+2)!}{(l-2)!}\int d^3K P_S(K)|\frac{3}{4}\int_0^{\eta_0}d\eta\,g(\eta)\,\Pi(K,\eta)\frac{j_l}{x^2} |^2,\label{Emode4}
\end{eqnarray}
and $\bar{\kappa}$ is the time average value of $\kappa(\eta)$
\begin{equation}\label{barkappa}
    \bar{\kappa}=\frac{1}{\eta_0}\int_0^{\eta_0}d\eta\,
    \bar{\kappa}\simeq\frac{\sqrt{2}}{12\pi}\alpha\,G^F\,n_{\nu,0}\frac{n_{\nu,0}}{z_{lss}}\int_0^{z_{lss}}dz\frac{(1+z)^3}{H(z)}\sim 0.16,
\end{equation}
where $z_{lss}\sim 1100$ indicates redshift at the last scattering surface, $n_{\nu,0}=\sum_j(\nu+\bar{\nu})\sim 340/cm^3$ is the total CNB number density of all flavor neutrinos at present time, and
$H(z)$ is the Hubble parameter. The term with $\bar{\kappa}$ shows that
the B-mode power spectrum can be generated  not only by the Compton scattering
in the case of tensor mode perturbations \cite{zal}, but also by the
CMB-CNB interaction in the case of scalar mode perturbation.

\section{Power Spectrum of Tensor Perturbation Modes}
The method of analysis used in previous section for the scalar perturbation can be utilized for the tensor perturbation. Bur the analysis is more complicate than that for the scalar perturbation because there are two independent polarized states denoted by $+ $ and
$ \times $ for each Fourier mode. It is convenient to use two linear combinations
\begin{equation}\label{xit}
     \xi^{1}= (\xi^{+}-i\xi^{\times})/2,\,\,\,\,\,\,\,\xi^{2}=(\xi^{+}+i\xi^{\times})/2,
\end{equation}
 where $\xi$'s are independent random variables and have the following statistical properties
\begin{eqnarray}
\langle \xi^{1*}(\vec{K_{1}})\xi^{1}(\vec{K_{2}})\rangle=\langle \xi^{2*}(\vec{K_{1}})\xi^{2}(\vec{K_{2}})\rangle=\frac{P_{h}(K)}{2}\delta(\vec{K}_{1}-\vec{K}_{2})~~~~,~~~~\langle \xi^{1*}(\vec{K_{1}})\xi^{2}(\vec{K_{2}})\rangle =0
\end{eqnarray}
where $P_{h}(K)$ is the primordial power spectrum of the gravitational wave.
Also we define
\begin{equation}\label{delta-pt}
    \tilde{\Delta}_{P^{\pm}}^{+(\times)}(\tau, \hat{n},\vec{K})= (\tilde{\Delta}_{Q}^{+(\times)}\pm i\tilde{\Delta}_{U}^{+(\times)})(\tau, \hat{n},\vec{K}).
\end{equation}
In the absence of CMB-CNB interaction, the polarizations generated by the gravitational wave  satisfy the following Boltzman equations \cite{crittenden93,polnarev}:
\begin{eqnarray}
\dot{\tilde{\Delta}}_{P^{\pm}}^{+(\times)}+ik\mu\tilde{\Delta}_{P^{\pm}}^{+(\times)}&=&-\dot{\tau}[\tilde{\Delta}_{P^{\pm}}^{+(\times)}+\psi(1\pm i)].
\end{eqnarray}
where $\dot{\tau}\equiv\frac{d\tau}{d\eta}$ and $ \eta $ is conformal time, the scaling factor $ a(\eta_{0}) $ at present time $ \eta_{0} $ is unity, and
\begin{eqnarray}
\psi\equiv [\frac{1}{10}\tilde{\Delta}_{T0}^{(T)}+\frac{1}{7}\tilde{\Delta}_{T2}^{(T)}+\frac{3}{7}\tilde{\Delta}_{T4}^{(T)}-\frac{3}{5}\tilde{\Delta}_{Q0}^{\pm(T)}+\frac{6}{7}\tilde{\Delta}_{Q2}^{\pm(T)}-\frac{3}{70}\tilde{\Delta}_{Q4}^{\pm(T)}].
\end{eqnarray}
In the presence of CMB-CNB interaction, the Boltzmann equations for polarizations generated by gravitational wave are modified
\begin{eqnarray}
\dot{\tilde{\Delta}}_{P^{\pm}}^{+(\times)}+ik\mu\tilde{\Delta}_{P^{\pm}}^{+(\times)}&=&-\dot{\tau}[\tilde{\Delta}_{P^{\pm}}^{+(\times)}+\psi(1\pm i)]\mp i a(\eta) \dot{\kappa}_{\pm}\tilde{\Delta}_{P^{\pm}}^{+(\times)}.
\end{eqnarray}
 The above equation has the following solution \cite{zal}
 \begin{equation}\label{deltapm}
    \tilde{\Delta}_{P^{\pm}}^{+(\times)}(\kappa ,\eta)=\int d\eta e^{ix \mu \pm i \kappa \pm i\frac{\pi}{4}} S_{P}(k,\eta)
 \end{equation}
where $ S_{P}(k,\eta)= \sqrt{2}\dot{\tau}(e^{-\tau })\psi$. For the tensor perturbation, the evolution equations take their simplest form after the coordinate transformation \cite{kosowsky96,polnarev},
 \begin{eqnarray}
  \Delta_Q^+ &=& (1+\mu^2)\cos(2\phi)\tilde{\Delta}_Q^+,\,\,\,\,\,\,\,\Delta_Q^\times = (1+\mu^2)\sin(2\phi)\tilde{\Delta}_Q^\times  \\
   \Delta_U^+ &=& -2\mu\sin(2\phi)\tilde{\Delta}_I^+,\,\,\,\,\,\,\,\,\,\,\,\,\,\,\,\,\,\,\Delta_U^\times = 2\mu\cos(2\phi)\tilde{\Delta}_U^\times .
 \end{eqnarray}
Tensor perturbations can be decomposed to the general form
\begin{eqnarray}
(\Delta_{Q}^{(T)}\pm i\Delta_{U}^{(T)})(\tau, \hat{n},\vec{K})&=&(\Delta_{Q}^{+}\xi^{+}+\Delta_{Q}^{\times}\xi^\times)(\tau, \hat{n},\vec{K})\pm i(\Delta_{U}^{+}\xi^{+}+\Delta_{U}^{\times}\xi^{\times})(\tau, \hat{n},\vec{K})\nonumber\\
(\Delta_{Q}^{(T)}\pm i\Delta_{U}^{(T)})(\tau, \hat{n},\vec{K})&=&  (1+\mu^{2})\cos 2\phi [\xi^{1}+\xi^{2}]\tilde{\Delta}_{Q}^{+}+(1+\mu^{2})\sin 2\phi [\xi^{1}-\xi^{2}]i \tilde{\Delta}_{Q}^{\times} \nonumber\\
&\pm &  i(2\mu)(-\sin 2\phi [\xi^{1}+\xi^{2}]\tilde{\Delta}_{U}^{+}+(1+\mu^{2})\cos 2\phi [\xi^{1}-\xi^{2}]i \Delta_{U}^{\times}).
\label{123}
\end{eqnarray}
Substituting  $\tilde{\Delta}_{Q}=\frac{1}{2}(\tilde{\Delta}_{P}^{+}+\tilde{\Delta}_{P}^{-})$ and $\tilde{\Delta}_{U}= (\frac{-i}{2})(\tilde{\Delta}_{P}^{+}- \tilde{\Delta}_{P}^{-})$ into Eqs.~(\ref{123}), we obtain
\begin{eqnarray}
(\Delta_{Q}^{(T)}\pm i\Delta_{U}^{(T)})(\tau, \hat{n},\vec{K})&=&  (1+\mu^{2})\xi^{1}e^{2i\phi}(\tilde{\Delta}_{P}^{+(T)}+\tilde{\Delta}_{P}^{-(T)}) +(1+\mu^{2})\xi^{2} e^{-2\phi} (\tilde{\Delta}_{P}^{+(T)}+\tilde{\Delta}_{P}^{-(T)})\nonumber\\
&\pm &  (-i)(2\mu)(\tilde{\Delta}_{P}^{+(T)}-\tilde{\Delta}_{P}^{-(T)})  [-\xi^{1}e^{2i\phi}-\xi^{2}e^{-2i\phi}].
\end{eqnarray}
Using $ 2 \mu = \frac{1}{2}[(1+\mu)^{2}-(1-\mu)^{2}]$ and $ 1+\mu^{2} = \frac{1}{2}[(1+\mu)^{2}+(1-\mu)^{2}]$ and then similarly to the scalar perturbation, we integrate along the line of sight
  \begin{eqnarray}
(\Delta_{Q}^{(T)}+ i\Delta_{U}^{(T)})(\tau, \hat{n},\vec{K})&=& \frac{1}{2}[(1+\mu)^{2}(1+2i)+(1-\mu)^{2}(1-2i)] [\xi^{1}e^{ 2i\phi }+\xi^{2}e^{ -2i\phi }]\tilde{\Delta}_{p}^{+}\nonumber\\&+&\frac{1}{2}[(1+\mu)^{2}(1-2i)+(1-\mu)^{2}(1+2i)] [\xi^{1}e^{ 2i\phi }+\xi^{2}e^{ -2i\phi }]\tilde{\Delta}_{p}^{-}\nonumber\\
(\Delta_{Q}^{(T)}- i\Delta_{U}^{(T)})(\tau, \hat{n},\vec{K})&=& \frac{1}{2}[(1+\mu)^{2}(1-2i)+(1-\mu)^{2}(1+2i)] [\xi^{1}e^{ 2i\phi }+\xi^{2}e^{ -2i\phi }]\tilde{\Delta}_{p}^{+}\nonumber\\&+&\frac{1}{2}[(1+\mu)^{2}(1+2i)+(1-\mu)^{2}(1-2i)] [\xi^{1}e^{ 2i\phi }+\xi^{2}e^{ -2i\phi }]\tilde{\Delta}_{p}^{-}.
\end{eqnarray}
Using  the spin raising and lowering operators twice as following,
  \begin{eqnarray}
  \bar{\eth}^{2}[(1\pm\mu)^{2}(1-\mu^{2})e^{ix\mu}]&=&[-\hat{\varepsilon}(x)\pm i\hat{\beta}(x)][(1-\mu^{2})e^{ix\mu}]\nonumber\\
  \eth^{2}[(1\pm\mu)^{2}(1-\mu^{2})e^{ix\mu}]&=&[-\hat{\varepsilon}(x)\pm i\hat{\beta}(x)][(1-\mu^{2})e^{ix\mu}]
  \end{eqnarray}
  where $ \hat{\varepsilon}(x)=-12+x^{2}[1-\partial_{x}^{2}]-8x\partial_{x} $ and $ \hat{\beta}(x)=8x+2x^{2}\partial_{x}$, we can separate the CMB linearly polarizations in terms of E-modes $\Delta_{E}^{(T)}$ and B-modes $\Delta_{B}^{(T)}$
   \begin{eqnarray}
  \Delta_{E}^{(T)}(\hat{n})&\equiv & \frac{-1}{2}[\bar{\eth}^{2}\Delta_{P}^{+(T)}(\hat{n})+\eth^{2}\Delta_{P}^{-(T)}]\nonumber\\
   \Delta_{B}^{(T)}(\hat{n})&\equiv& \frac{i}{2}[\bar{\eth}^{2}\Delta_{P}^{+(T)}(\hat{n})-\eth^{2}\Delta_{P}^{-(T)}].
  \end{eqnarray}
As a result, we obtain
   \begin{eqnarray}
   \Delta_{E}^{(T)}(\eta,\hat{n},\vec{k})&=& \sqrt{2} [(1-\mu^{2})e^{2i\varphi}\xi^{1}(\vec{k})+(1-\mu^{2})e^{-2i\varphi}\xi^{2}(\vec{k})]\int_{0}^{\eta_{0}}d\eta e^{ix\mu} S_{P}^{T}(k,\eta)[\hat{\varepsilon}(x)\cos(\kappa(\eta)+ \frac{\pi}{4})]\nonumber\\
   \Delta_{B}^{(T)}(\eta,\hat{n},\vec{k})&=&\sqrt{2} [(1-\mu^{2})e^{2i\varphi}\xi^{1}(\vec{k})-(1-\mu^{2})e^{-2i\varphi}\xi^{2}(\vec{k})]\int_{0}^{\eta_{0}}d\eta e^{ix\mu} S_{P}^{T}(k,\eta)[\hat{\beta}(x)\sin(\kappa(\eta)+ \frac{\pi}{4})].\nonumber\\
     \end{eqnarray}
		
The E-mode power spectrum is given as
     \begin{eqnarray}
       C_{El}^{(T)}=(4\pi)^{2}\int k^{2}dk P_{h}(k) (\int_{0}^{\eta_{0}}d\eta e^{ix\mu} S_{P}^{(T)}(k,\eta)[\hat{\varepsilon}(x)\frac{\cos(\kappa(\eta)+ \frac{\pi}{4})}{\cos\frac{\pi}{4}}]\frac{j_{l}(x)}{x^{2}})^{2},\label{CET}
     \end{eqnarray}
which is the same as that in the standard scenario for CMB polarizations without considering the CNB-CMB interactions, except for an additional factor $(\frac{\cos(\kappa(\eta)+ \frac{\pi}{4})}{\cos\frac{\pi}{4}})^2$. This factor becomes unit if the CNB-CMB interaction ($\kappa\rightarrow0$) is neglected and the result of standard scenario for CMB polarizations
for the E-mode can be yielded.
Using similar method , we obtain the power spectrum of B-mode
     \begin{eqnarray}
       C_{Bl}^{(T)}=(4\pi)^{2}\int k^{2}dk P_{h}(k) (\int_{0}^{\eta_{0}}d\eta e^{ix\mu} S_{P}^{(T)}(k,\eta)[\hat{\beta}(x) \frac{\sin(\kappa(\eta)+ \frac{\pi}{4})}{\sin(\frac{\pi}{4})}]\frac{j_{l}(x)}{x^{2}})^{2}.\label{CBT}
     \end{eqnarray}
Similar to the E-mode power spectrum $C_{El}^{(T)}$,
the B-mode power spectrum $C_{Bl}^{(T)}$ is the same as a result of
the standard scenario for CMB polarizations, except an additional
factor $(\frac{\sin(\kappa(\eta)+ \frac{\pi}{4})}{\sin(\frac{\pi}{4})})^2$
which becomes unit when $\kappa\rightarrow0$.
The expressions for the power spectra of  E and B modes are different in $\kappa$-dependent terms, also in the $\hat{\varepsilon}$ and $\hat{\beta}$ operators, which is obtained after the angular integrals.

\begin{figure}
  \includegraphics[width=4in]{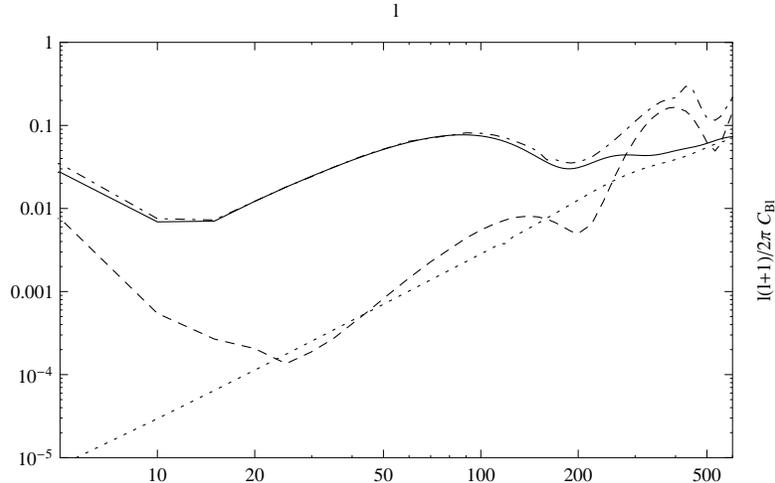}\\
  \caption{The B-mode power spectrum $C_{Bl}l(l+1)/2\pi$ is plotted in terms of $l$ and in unit $(\mu K)^2$. Solid curve is for the B-mode due to Compton scattering in the presence of tensor perturbations and lens effects. Dashed curve is for
CNB-CMB interaction and Compton scattering in the case of scalar perturbation.
Dotted curve shows lens effects, while dot-dashed curve indicates
the contribution of lens effects and Compton scattering in the presence of tensor perturbations plus CNB-CMB interactions.}\label{ClB-1}
\end{figure}

\section{ Conclusion.}
The generation of the B-mode spectrum of CMB photons in the cases of scalar and tensor
perturbations is calculated by using  Quantum
Boltzmann Equation for the density matrix or Stokes parameters of CMB photons, taking
into account both CMB-CNB and Compton scatterings.
As shown in Eqs.(\ref{CBS1}), (\ref{CET}) and (\ref{CBT}), the power spectra of
E-mode and B-mode of CMB polarization are modified in the presence of CNB-CMB interaction.  The most important point is that the B-mode is generalized by the
CNB-CMB interaction in the case of scalar perturbation.
The generated B-mode spectrum is approximately proportional to
$\propto(\alpha\, G^F\, H_0^{-1}\, \bar{n}_\nu)^2$, where $\bar{n}_\nu$ is the
average CNB number density from the last scattering up to present time.
As discussed in Ref.~\cite{r.mohammadi}, for the reason of neutrinos
being left-handed and their gauge-couplings being parity violated,
linearly polarized photons acquire their circular and
magnetic-like linearly polarizations by interacting with neutrinos.
It is important to mention that cosmic neutrinos can be either Dirac or Majorana types, the rate of B-mode generation $\dot{\kappa}_{\pm}$ for the Dirac neutrino case is about twice less than the rate for the Majorana neutrino case \cite{r.mohammadi}.

The B-mode power spectrum of CMB for different cases is plotted in Fig.(\ref{ClB-1}). This result shows that the contribution of CNB-CMB interaction have caused
an enhancement in $C_{Bl}$ for $l>300$, which could be used to show the evidence of CNB, since the CNB neutrinos have not yet been directly detected, due to their week interactions
and very low energy. To end this article, we would like to emphasize that the
$r$-parameter is usually calculated by
comparing B-mode  and E-mode power spectrum ${\rm r=P_T/P_S\propto C^T_{Bl}/C^S_{El}=C^{ob}_{Bl}/C^S_{El}}$ and assuming that the observed B-mode $C^{ob}_{Bl}$ is totally attributed to
the Compton scattering in the presence of tensor perturbations $C^T_{Bl}$. However our calculations show that other interactions like CNB-CMB interaction can have contribution to
the B-mode power spectrum in the presence of scalar mode $C_{Bl}^S$. For this situation the exact value of r-parameter is suppressed like  ${\rm r\propto C^T_{Bl}/C^S_{El}\propto (C^{ob}_{Bl}-C^S_{Bl})/C^S_{El}}$.

\section{Acknowledgements}
This  study  was  supported  by {\bf Iran National Science Foundation: INSF} also R. Mohammadi  and J. Khodagholizadeh  would like to thanks from {\it School of physics, Institute for research in fundamental sciences (IPM), Tehran, Iran}.



\begin{thebibliography}{99}
\bibitem{BICEP2}
  P.~A.~R.~Ade {\it et al.}  [BICEP2 Collaboration],
 ``BICEP2 I: Detection Of B-mode Polarization at Degree Angular Scales,''[arXiv:1403.3985 [astro-ph.CO]]
  \bibitem{Bovin}
  C.~Bonvin, R.~Durrer and R.~Maartens,
  ``Can primordial magnetic fields be the origin of the BICEP2 data?,''Phys.\ Rev.\ Lett.\  {\bf 112}, 191303 (2014).[arXiv:1403.6768 [astro-ph.CO]].
\bibitem{14}
C. Scoccola, D. Harari and S. Mollerach, Phys. Rev. D 70, 063003 (2004); L. Cam-
panelli, A. D. Dolgov, M. Giannotti and F. L. Villante, Astrophys. J. 616, 1 (2004);
A. Kosowsky, T. Kahniashvili, G. Lavrelashvili and B. Ratra, Phys. Rev. D 71, 043006
(2005); M. Giovannini, Phys. Rev. D 71, 021301 (2005); M. Giovannini and K. E. Kunze,
Phys. Rev. D 78, 023010 (2008); Phys. Rev. D 79, 063007 (2009); L. Pogosian,
A. P. S. Yadav, Y. -F. Ng and T. Vachaspati, Phys. Rev. D 84, 043530 (2011); M. Gio-
vannini, Phys. Rev. D 89, 061301 (2014); C. Bonvin, R. Durrer and R. Maartens,
arXiv:1403.6768 [astro-ph.CO].
\bibitem{Giovannini}
  M.~Giovannini,
  ``Faraday scaling and the Bicep2 observations,''Phys.\ Rev.\ D {\bf 90}, 041301 (2014).[arXiv:1404.3974 [astro-ph.CO]].
  \bibitem{Khodagholizadeh}

  J.~Khodagholizadeh, R.~Mohammadi and S.~S.~Xue,
  ``Photon-neutrino scattering and the B-mode spectrum of CMB photons,''[arXiv:1406.6213 [astro-ph.CO]].

\bibitem{dust}
  R.~Adam {\it et al.} [Planck Collaboration],
  arXiv:1409.5738 [astro-ph.CO].
  \bibitem{keck}
  P.~A.~R.~Ade {\it et al.} [BICEP2 and Keck Array Collaborations],
  arXiv:1510.09217 [astro-ph.CO].
  \bibitem{nagata}
  T.~Namikawa, D.~Yamauchi, B.~Sherwin and R.~Nagata,
  arXiv:1511.04653 [astro-ph.CO].

\bibitem{Dicus}  D.~A.~Dicus and W.~W.~Repko,
 Phys.\ Rev.\ D {\bf 48}, 5106 (1993).[hep-ph/9305284].

\bibitem{r.mohammadi}
  R.~Mohammadi and S.~S.~Xue,Phys.\ Lett.\ B {\bf 731}, 272 (2014).[arXiv:1312.3862 [hep-ph]]; R. Mohammadi, Eur.\ Phys.\ J.\ C {\bf 74}, no. 10, 3102 (2014), [arXiv:astro-ph.CO/1312.2199v1];
  P.~Bakhti, R.~Mohammadi and S.~-S.~Xue,  [arXiv:1403.7327 [hep-ph]].
\bibitem{cosowsky1994}
A.~Kosowsky, Annals Phys. {\bf246}, 49-85 (1996),
[arXiv:astro-ph/9501045].
  arXiv:1403.7327 [hep-ph].
\bibitem{zal} M. Zaldarriaga and U. Seljak,
Phys. Rev.  {\bf D55}, 1830 (1997), [astro-ph/9609170];  M.~Zaldarriaga, D.~N.~Spergel and U.~Seljak,
  Astrophys.\ J.\  {\bf 488}, 1 (1997)
  [astro-ph/9702157].
\bibitem{hu} W. Hu and M. J. White, �A CMB Polarization Primer,� New Astron. {\bf2}, 323 (1997),
[arXiv:astro-ph/9706147].

\bibitem{chandra} S. Chandrasekhar, ``Radiative Transfer'', Dover,
New York, 1960.
\bibitem{kaiser} N. Kaiser, Mon. Not. R. Astron. Soc. {\bf 202}, 1169 (1983).

\bibitem{sz} U. Seljak, and M. Zaldarriaga, Astrophys. J. {\bf 469},
437 (1996).
\bibitem{letter} U. Seljak, and M. Zaldarriaga,  Report no.
astro-ph/9609169, 1996 (unpublished).
\bibitem{zh95}
M. Zaldarriaga and D. Harari, Phys. Rev. D {\bf 52}, 3276 (1995).


\bibitem{uros} U. Seljak, Report no. astro-ph/9608131, 1996 (unpublished).
\bibitem{crittenden93}
R. Crittenden, R. L. Davis, and P. J. Steinhardt,
Astrophys. J. Lett. {\bf 417}, L13 (1993).
\bibitem{polnarev} A. G. Polnarev, Sov. Astron. {\bf 29}, 607 (1985).

\bibitem{kosowsky96} A. Kosowsky, Ann. Phys. {\bf 246}, 49 (1996).





\end{thebibliography}
\end{document}